# Monolayer H-Si-P Semiconductors: Structural stability, electronic structure, optical properties, and Prospects for photocatalytic water splitting

Xiaoqin Shu[1], Jiahe Lin[2], Hong Zhang[3*]

[1] School of Mathematics and Physics , Leshan Normal College, Leshan, 614000, China

[2] School of science, Jimei University, Fujian,361021,China

[3] College of Physical Science and Technology, Sichuan University, Chengdu, 610065, China

Group IV and V monolayers are the promising state-of-the-art 2D materials owing to their high carrier mobility, tunable bandgaps, and optical linear dichroism along with outstanding electronic and thermoelectric properties. Furthermore, recent studies reveal the stability of free-standing 2D monolayers by hydrogenation. Inspired by this, we systematically predict and investigate the structure and properties of various hydrogen saturated silicon phosphide (H-Si-P) monolayers, based on first-principles calculations. According to the results, H-Si-P monolayers belong to indirect bandgap semiconductors with a highly stable structure. Their bandgaps and band edge positions assessed using accurate hybrid functional are shown to be effectively adjusted by applying a biaxial strain. Furthermore, the absorption spectra of these monolayers, simulated in the context of time-dependent density functional theory, exhibit their excellent potential for solar energy conversion and visible-light-driven photocatalytic water splitting. In this respect, this work provides valuable guidance for finding more 2D semiconductors and nanostructures for nanoelectronic and optoelectronic applications, as well as for photocatalytic water splitting.

Key words: 2D monolayers, 2D semiconductors and nanostructures, first-principles calculations; photocatalytic water splitting.

---

[*] Corresponding author.

*E-mail address*: hongzhang@scu.edu.cn（H. Zhang ）

1. Introduction

Two-dimensional (2D) materials have attracted increasing interest with respect to their integration into optoelectronic devices. Among them, graphene, $MoS_2$, $MoSe_2$ and so on [1-8], due to the tunable band gap that is the basis for broadband photoresponse. However, the in-plane isotropic structure of these materials impedes their application in polarization-sensitive photodetectors. In this respect, since 2014, tremendous attention has been paid to a novel 2D layered semiconductor material, called black phosphorus (BP). Its in-plane anisotropic physical properties along zigzag and armchair directions along with prominent carrier mobility and thickness-dependent direct band gap allowed BP to be considered as promising not only in polarization-sensitive photodetectors, but also in transistors, photonic and optoelectronic appliances, sensors, batteries, catalysis, and many other applications [9-17]. Meanwhile, the extensive development of BP was hampered by its instability under ambient conditions, as well as by the lack of techniques for producing large-area and high-quality 2D nanofilms[18,19].

In this regard, attempts were made to discover analogues with improved characteristics, leading to group IV-V 2D semiconductors, such as GeP [20] and GeAs [21] materials, whose high in-plane anisotropic properties have become a hot topic of modern research. Among these, particular attention is drawn to orthorhombic silicon phosphides (o-SiP, mm2 point group), for which various theoretical calculations predict a widely tunable band gap (from 1.69 to 2.59 eV) and high carrier mobility similar to that of black phosphorus [22]. However, there is still a few experimental works that could bring new information on the optical and electronic properties, especially on the in-plane anisotropic characteristics of o-SiP. One rare example is a study of Li *et al.*, who have experimentally confirmed that o-SiP is an excellent optoelectronic 2D material with a large band gap (1.71 eV), high mobility (2.034 × $10^3$ $cm^2 \cdot V^{-1} \cdot s^{-1}$), and fast photoresponse.[23]. Nevertheless, the in-plane anisotropy of o-SiP, being of importance for state-of-the-art devices, such as thin-film polarizers, polarization sensors, and plasmonic devices, has still remained beyond of the scope of many researchers.

In this work, we design three hydrogen saturated SiP monolayers, called H-Si-P. Depending on the hydrogen position, they are HPSi, HSiP, and HSiPbp structures, respectively. HPSi means that hydrogen atoms are added to phosphorus atoms in the upward direction. HSiP refers to a structure where hydrogen atoms are attached to silicon atoms in the upward direction. HSiPbp indicates that hydrogen atoms are linked to silicon atoms in both the upward and downward directions. The HPSi and HSiP monolayers possess a space-group symmetry $P3M1$, whereas HSipbp monolayer is described by a $PMN21$ space group. Special attention is paid to a theoretical study of the optical and electronic properties of these monolayers simulated through density functional theory (DFT) . In particular, their phonon spectra and cohesive energies are calculated to prove a highly stable structure of these compounds. The investigation of band gaps and band edge positions of monolayer H-Si-Ps exhibits their semiconductor properties, which even can be tuned by applying the mechanical biaxial strains. The optical absorption spectra of the monolayer compounds reveal an increase in the potential efficiency of solar energy conversion and water splitting, proving that these nanostructures are promising for water splitting in a visible-light region.

## 2. Computational details

For periodic H-Si-P monolayer, the optical and electronic properties were simulated in the context of the ab initio density functional theory (DFT) using the CASTEP package [24]. During calculations, the norm-conserving pseudopotentials and the plane wave energy cutoff of 720 eV were employed to relax the structure models along with the band structures via the generalized gradient approximation (GGA) expressed by the Perdew-Burke-Ernzerhof (PBE) functional [25]. The structures, including the primitive cells of monolayer H-Si-P, were exposed to relaxation until the forces were smaller than 0.01 eV/Å and the energy tolerances were less than $1\times10^{-7}$ eV per atom. A vacuum layer of 23 Å was used to avoid interactions between neighboring layers. The k-point sampling of the Brillouin zone with 6×6×1 for monolayer HPSi and HSiP, with 8×8×1 for HSiPbp was performed by adopting the Monkhorst-Pack scheme[26]. The phonon dispersions of monolayer H-Si-P were calculated by the linear response method [27]. The band structures, density of states, and optical properties were evaluated by the Heyd-Scuseria-Ernzerhof screened hybrid functional (HSE06) [28].

For monolayer nanostructures, the collective plasmonic excitations were simulated in the direction parallel to the nanostructure plane using the time-dependent density functional theory (TDDFT) on the basis of OCTOPUS real-space TDDFT code [29]. The dangling bonds at the edges were passivated by hydrogen atoms using the Troullier–Martins pseudopotentials [30] to elucidate the impact of different edge states and dimensional confinement on the collective plasmonic excitations of H-Si-P structures. Reparametrized PBE (Perdew, Burke and Ernzerhof)[25] for the van der Waals interactions for the exchange potential and correlation potential used in both the ground state and excited-state calculations. To obtain the linear optical absorption spectra of the studied structures, their initial states were excited by providing a small momentum $E(t) = E_{kick}\delta(t)$ to the electrons at assuming the time-propagating Kohn–Sham wave functions. The information of the excitations was deduced from the dipole-strength function, and the excitation spectrum was obtained from the Fourier transformation of the dipole strength. The simulation zone was defined by assigning a sphere centered around each atom with a radius of 7 Å and a uniform mesh grid of 0.3 Å. In a real-time propagation, the Kohn–Sham wave function evolved for typically 6000 steps with a time step of 0.003 $\hbar$ / eV. The atoms were fixed in the XY plane of the Cartesian coordinate system. The zigzag edge of the nanostructure was assumed to be perpendicular to the X-axis, while the armchair edge was parallel to the X-axis.

## 3. Results and discussion

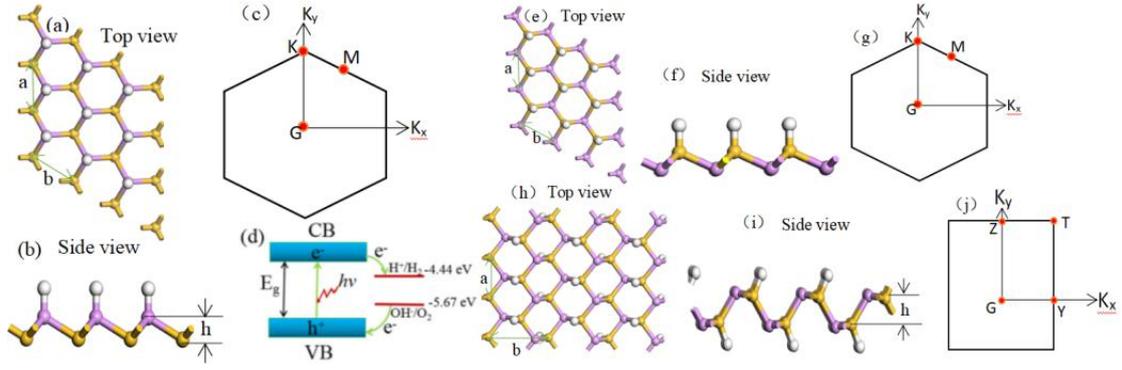

FIG. 1. (a), (e) and (h) Top views of the monolayer HPSi, HSiP, and HSiPbp structures, respectively, marked with their structural parameters. (b), (f) and (i) Side views of the monolayer HPSi, HSiP, HSiPbp structures, respectively. The gold, purple, and white spheres stand for silicon, phosphorus, and hydrogen atoms, respectively. (c), (g) and (j) Brillouin zones and main high symmetric points of the monolayer HPSi, HSiP, and HSiPbp structures, respectively. (d) Illustration of photocatalytic water splitting. The successful water splitting using a single photocatalyst is achieved through a properly aligned bandgap of the semiconductor with respect to the redox potentials of water -4.44 eV and -5.67 eV, respectively.

Three different structures, so-called HPSi, HSiP, and HSiPbp, were simulated. Their full structural relaxations enabled one to obtain the structural parameters, including lattice constants, bond length, layer height, and bond angles as shown in Figure 1. The corresponding values of the parameters evaluated for the above structures are summarized in Table I. In order to study the stability of H-Si-P monolayers, their cohesive energies along with the phonon dispersions were calculated using the equations below:

$$E_{coh} = \frac{E_{tot} - E_{Si} - E_P - E_H}{3} \quad (1)$$

$$E_{coh} = \frac{E_{tot} - 2E_{Si} - 2E_P - 2E_H}{6} \quad (2)$$

Equation (1) was used for HPSi and HSiP structures, and Eq. (2) was applied for HSiPbp. Here, $E_{coh}$ is the total energy of H-Si-P monolayer and $E_{Si}$, $E_P$, $E_H$ are the ground state energies of Si, P, and H free atoms, respectively, calculated in neglecting the interactions between the neighboring atoms in a cubic cell with a lattice constant of 20 Å. As seen in Table 1, the cohesive energies of HPSi, HSiP and HSiPbp are found to be -4.498, -4.751, and -4.740 eV, respectively. The high enough cohesive energies of the three monolayer structures evidence the high stability of H-Si-P monolayers.

The properties of phonons can be described using a harmonic approximation based on the knowledge of just one fundamental quantity, the force constants matrix:

$$D_{\mu\nu}(R - R') = \left.\frac{\partial^2 E}{\partial u_\mu(R)\, \partial u_\nu(R)}\right|_{\nu=0} \quad (3)$$

Here $\mu$ refers to the displacement of a given atom and $E$ is the total energy in the harmonic approximation. This force constants matrix (or Hessian matrix) can also be represented in reciprocal space and the result is commonly referred to as the dynamical matrix:

$$D_{\mu\nu}(q) = \frac{1}{N_R}\sum_R D_{\mu\nu}(R)\exp(-iqR) \quad (4)$$

Classical equations of motion can be written in the language of dynamical matrices,

as an eigenvalue problem. Each atomic displacement is described in the form of plane waves:

$$u(R,t) = \varepsilon e^{i(qR-\omega(q)t)} \quad (5)$$

where the polarization vector of each mode, $\varepsilon$, is an eigenvector with the dimension of 3N of the eigenvalue problem:

$$M\omega(q)^2\varepsilon = D(q)\varepsilon \quad (6)$$

The dependence of the frequency, $\omega$, on the wave vector is known as the phonon dispersion.

Figure 2 displays the phonon band dispersions of HPSi, HSiP, and HSiPbp structures along the high symmetric points of their Brillouin zones with respect to Figures 1(g)-1(i). Here, the stability of the calculated phonon dispersions at a lack of soft modes and the linear dispersion relation of the acoustic branch around the G point indicates the outstanding kinetic stability of the simulated monolayer H-Si-P structures.

TABLE I. Optimized geometries and cohesive energies of H-Si-P (HPSi, HSiP, HSiPbp) monolayers, obtained using DFT with a PBE exchange-correlation functional.

| Structure | Space group | Cohesive energy (eV/atom) | Lattice constant (Å) a | b | Bond length (Å) l | Layer height(Å) h |
|---|---|---|---|---|---|---|
| HPSi | *P3M1* | -4.498 | 3.5469 | 3.5469 | 2.323 | 1.09 |
| HSiP | *P3M1* | -4.751 | 3.5427 | 3.5469 | 2.269 | 1.09 |
| HSiPbp | *PMN21* | -4.740 | 3.5345 | 5.5693 | 2.261 | 1.38 |

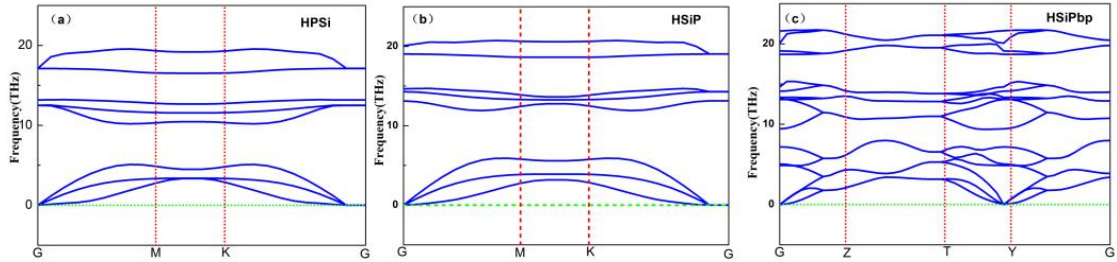

FIG. 2. (a)-(c) Phonon band dispersions of HPSi, HSiP, and HSiPbp structures.

Figure 3 depicts the band structures of monolayer H-Si-P systems, calculated using the HSE06 hybrid functional. Their bandgaps were found to be 2.74, 3.325, and 3.749 eV for HPSi, HSiP, and HSiPbp, respectively. Such the wide band gaps of the studied 2D materials allow one to refer them to 2D semiconductors that are suitable for applications in high-power electronic devices, field emission appliances, and optoelectronic tools operating under UV or visible light conditions. Furthermore, the bandgaps exhibit the indirect behavior. For HPSi, the VB maximum (VBM) lies at K point, and the CB minimum (CBM) occurs along the G-M direction. For HSiP, the VBM intersects the G point, and the CBM crosses over the M point. For HSiPbp, the VBM arises along the Y-G direction, and the CBM lies at G point.

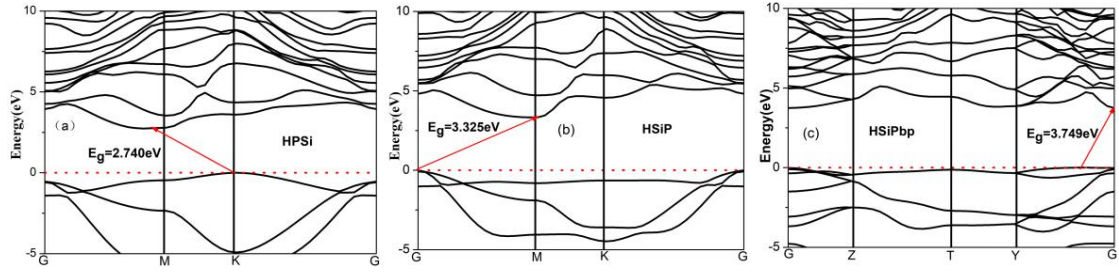

FIG. 3 (a)-(c) Band structures of monolayer HPSi, HSiP, and HSiPbp, respectively. The VBM is set to zero by the red dash lines.

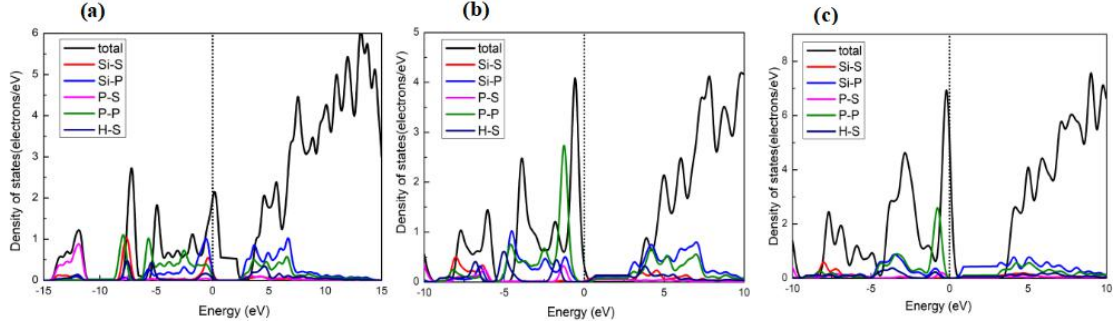

FIG. 4. (a)–(c) The total densities of states (TDOS) and partial densities of states (PDOS) of HPSi, HSiP, and HSiPbp , respectively.

In order to understand the contribution of different orbitals to the electronic states and the bonding characteristics of monolayer H-Si-P, the total densities of states (TDOS) and partial densities of states (PDOS) were calculated for monolayer H-Si-P. Figure 4 displays the total densities of states (TDOS) and partial densities of states (PDOS) calculated for monolayer H-Si-P As seen in Figure 4, TDOS of all monolayer compounds, considered in this work, exhibit multiple van Hove singularities over the entire energy range, which is consistent with the 2D nature of a monolayer SiP material. In turn, the PDOS of monolayer compounds reveal the contributions from both the $s$ and $p$ orbitals of Si and P near the Fermi level. Obviously, the effects from the $p$ orbitals of Si and P to the TDOS are much more pronounced than those from the $s$ orbitals. Such a predominance of the $p$ orbitals is due to the $sp^3$-like bonding of P atoms and $sp^2$-like bonding of group V forming the monolayer H-Si-P, and this feature is always observed in group IV diamond-like structures and monolayer honeycomb systems of group III elements. A thorough analysis of PDOS plots reveals that the state closest to the VBM of HPSi originates from the $p$ orbitals of Si atoms, whereas the states closest to the VBM of HSiP and HSiPbp arise from the p orbitals of P atoms. As for the states closest to the CBM, these refer to the $p$ orbitals of Si atoms. Additionally, the specific distribution of VBMs and CBMs in the monolayer compounds is beneficial for the separation of photogenerated electron-hole pairs, thus reducing their recombination and increasing the photocatalytic activity [31].

To determine the alignment of the CBM and VBM energies, the work functions of monolayer H-Si-P were calculated by using the HSE06 functional. Figure 5 displays the

CBM and VBM energy levels with the redox potentials of water splitting. To make a semiconductor promising for water splitting, both the reduction and oxidation potentials must be located inside the bandgap. As shown in Figure 5, their values are found to be $V_{H^+/H^2} = 4.44$ eV and $V_{OH^-/O_2} = 5.67$ eV, respectively, which are obviously within the band gaps of the studied materials, thereby revealing the energetically favorable redox process for these. In this respect, these monolayer materials can be deemed to be the candidates for photocatalytic water splitting. However, since the bandgaps of HSiP and HSiPbp monolayer compounds are within the UV light energy range, this impedes their application for visible-light-induced photocatalytic splitting water. Moreover, even though the theoretical calculations predict the suitability of monolayer H-Si-P systems for photocatalytic water splitting in the vacuum, the situation may change when placing them in a liquid water environment. Therefore, further investigations are necessary to understand the band structure behavior of these 2D materials in water.

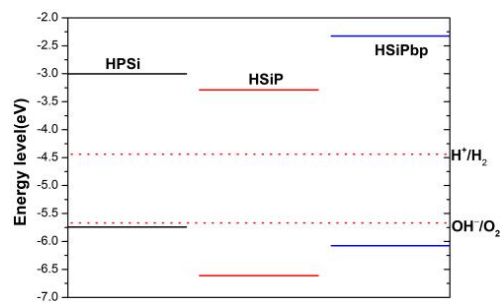

FIG. 5. Band edge positions of monolayer HPSi, HSiP, and HSiPbp relative to the vacuum level at a zero strain, calculated using the HSE06 functional. The standard redox potentials for water splitting are shown for comparison.

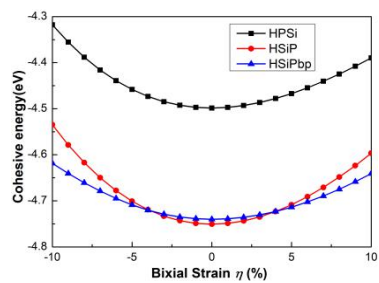

FIG. 6. Cohesive energies for HPSi, HSiP, and HSiPbp vs. biaxial strain.

Strain is inevitable in real systems due to the synthetic and application environments. Many experimental and theoretical studies [32-36] have shown that applying mechanical strain to the sample is a powerful method for modulating its band structure and the optical properties. In this respect, monolayer H-Si-P systems were subjected to mechanical tensile and compressive biaxial strains to monitor the evolution of their band structures and optical properties. The mechanical biaxial strain was simulated by freezing one of the lattice constants, as follows: $\eta = (a - a_0)/a_0$, where $a_0$ is the optimized lattice constant and $a$ is the lattice parameter along the strain direction. The positive and negative values of $\eta$ were attributed to tensile and compressive strain, respectively. Figure 6 displays the cohesive energy as a function of strain $\eta$ for the H-Si-P systems within a strain range from -10% to 10% at a spacing of 1%. According to these plots, the structures under consideration remain highly stable under the applied loads, even though the stability of stretched and compressed monolayers is weaker than that before strain. Furthermore, the HSiP and HSiPbp systems are obviously more stable than HPSi.

Figure 7 depicts the bandgaps versus strain $\eta$ for the H-Si-P structures within the same strain range, as in Figure 6. As seen in Figure 7(a), the band gaps of all the studied

H-Si-P monolayers exhibit the identical trend, increasing in a quasi-linear manner to a certain $\eta$ value and then linearly decreasing with increasing $\eta$. Special attention is drawn to the band gap variation for a monolayer HSiPbp (shown with a blue line in Figure. 7(a)), where two linear tendencies (a rise followed by a drop) of band gap values are clearly observed. This can be interpreted, as follows. At $\eta > -1\%$, there is an increase in the distance between the Si and P atoms, which makes the overlap integral of the wave functions between them decrease, leading to a decrease of their bandgaps. At $\eta < -1\%$, the distance between the neighboring Si atoms in the same sublayer becomes so close that the overlap integral of the wave functions for the inner electrons of these Si atoms increases, causing a decrease in their bandgaps with $|\eta|$ increasing. It is thus clear that the bandgaps of the monolayer H-Si-P have a linear response within a certain biaxial tensile strain range, which means the prospects of their application in mechanical sensors. For HPSi, the bandgaps within a range of -10% < $\eta$ < 10% are found to vary from 1.628 to 2.056 eV, respectively, thus meeting requirements for the photocatalytic water splitting under visible light. For HSiP, the bandgaps within a range of -10% < $\eta$ < -3 % are found to vary from 1.866 to 3.136 eV, respectively, and when $\eta$ = 8%, 9% and 10%, their bandgaps are 3.111, 2.857 and 2.613eV, respectively. For HSiPbp, the bandgaps within a range of -10% < $\eta$ < -5 % are found to vary from 1.862 to 3.129 eV, respectively and bandgaps within a range of 5 % < $\eta$ < 10 % are found to vary from 3.117 to 2.556eV, respectively. In this respect, the bandgaps of all presented 2D materials cover the visible light range, meaning that these semiconductors exhibit the high light utilization rate.

The effect of strain on the band edge positions of monolayer compounds was elucidated by determining the corresponding CBM and VBM from the relaxed configurations by calculating the work functions. Figure 7(b) displays the band edge positions of HPSi, HSiP, and HSiPbp undergoing biaxial strain. Therefore, Figure 7 is shown to provide useful guidance for tuning the bandgaps along with CBM and VBM levels of monolayer compounds to maximize the solar energy conversion efficiency. For this reason, data presented in Figure 7 were further used to calculate the absorption spectra of monolayer H-Si-P to investigate their sunlight utilization.

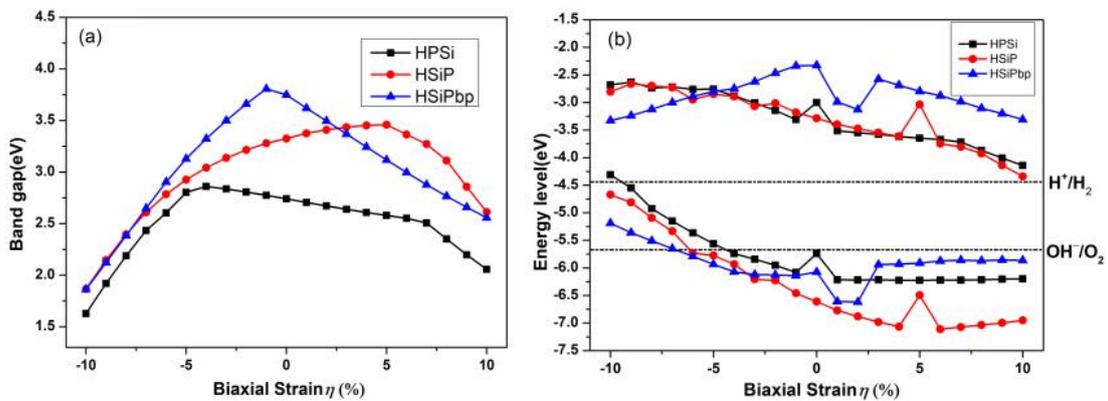

FIG. 7. (a) Bandgaps and (b) band edge positions of HPSi, HSiP, and HSiPbp vs. biaxial strain.

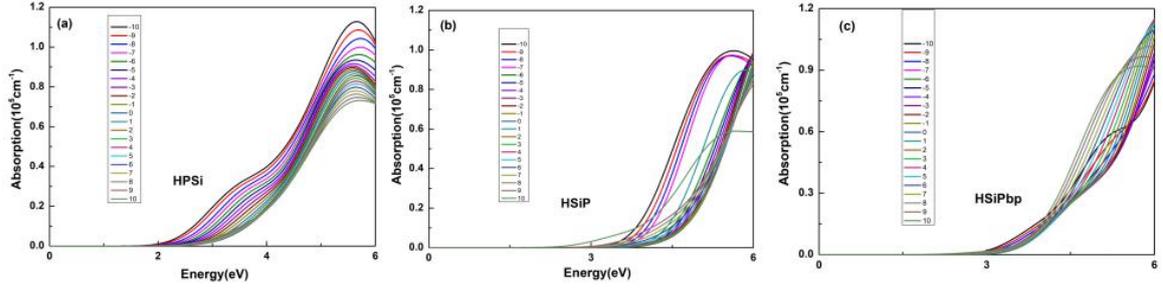

FIG. 8.Evolution of absorption spectra of (a) HPSi, (b) HSiP, and (c) HSiPbp systems with biaxial strain below 6eV.

The absorption spectra of 2D monolayer structures were calculated in assuming the in-plane polarized light. First, the frequency-dependent dielectric function

$\varepsilon(\omega) = \varepsilon_1(\omega) + i\varepsilon_2(\omega)$ was found. Then the absorption coefficient was evaluated as a function of photon energy according to the following expression [37]:

$$\alpha(E) = \frac{4\pi e}{hc}\left[\frac{(\varepsilon_1^2 + \varepsilon_2^2)^{1/2} - \varepsilon_1}{2}\right]^{1/2} \quad (7)$$

Figure 8 shows the absorption spectra of HPSi, HSiP, and HSiPbp under biaxial strain, simulated within a range of below 6eV. No absorption resonance peaks arise within a visible (3.3–1.7 eV) range. As is known, a periodic 2D graphene structure manifests itself by a resonance peak around 5.0 eV in its absorption spectrum [38-40]. However, in the case of 2D graphene nanostrctures, the boundary effect leads to the absorption peak splitting in the low energy region [41,42]. In this respect, one can assume the same situation for the monolayer H-Si-P nanostructures under consideration.

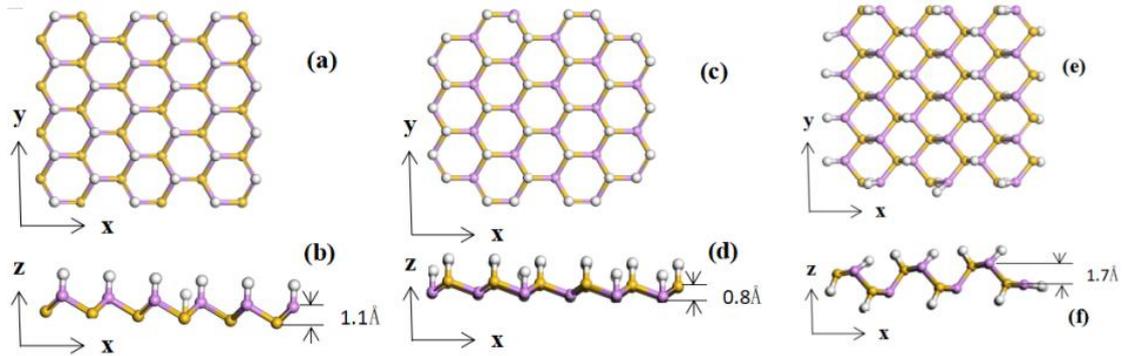

FIG. 9 Top and side views of the monolayer (a) and (b) for HPSi,(c) and (d) for HSiP, (e) and (f) for HSiPbp nanostructures, respectively, and their structural parameters. The HPSi nanostructure is composed of 27 silicon, 27 phosphorus and 27 hydrogen atoms. The HSiP nanostructure is composed of 25 silicon, 25 phosphorus and 34 hydrogen atoms in HSiP nanostructure. The HSiPbp nanostructure is composed of 27 silicon, 27phosphorus and 37 hydrogen atoms. In all structures, the dangling bonds at the edges are passivated by hydrogen atoms.

In order to confirm or deny this suggestion, the absorption peculiarities were calculated for the monolayer H-Si-P nanostructures below 9 eV, shown in Figure 10. For this, the collective plasmonic excitations in the monolayer H-Si-P nanostructures were investigated using the time-dependent density functional theory (TDDFT) [28].The linear optical absorption spectra of the systems were obtained following a scheme proposed by Yabana and Bertsch [43], where their excitation frequencies were determined from the momentum ($\kappa$) of the electron. The latter was achieved by transforming the ground-state wave functions propagating in some (finite) time:

$$\psi_i(\mathbf{r},\delta t) = e^{i\kappa z}\psi_i(\mathbf{r},0) \qquad (8)$$

The spectra could then be calculated from the expression of the dipole strength function presented below:

$$S(\omega) = \frac{2\omega}{\pi}\Im\alpha(\omega), \qquad (9)$$

where $\alpha(\omega)$ is the dynamical polarizability described by the Fourier transform of the dipole moment of the system $d(t)$ as

$$\alpha(\omega) = \frac{1}{\kappa}\int dt e^{i\omega t}[d(t)-d(0)] \qquad (10)$$

Using this definition, the Thomas-Reiche-Kuhn sum rule for the number of electrons ($N$) is given by the integral:

$$N = \int d\omega S(\omega) \qquad (11)$$

Figure 10 shows the optical absorption spectra of monolayer H-Si-P systems, simulated using Eq. (5). In comparison with spectra in Figure 8, those in Figure 10 were extended toward the infrared (below 1.7 eV) region, where it observes the absorption splitting for all the studied H-Si-P systems by analogy with a monolayer graphene nanostructure, which is due to the finite size effects[39,40]. Furthermore, the absorption spectrum of the graphene nanostructure is also known to depend on the edge (zigzag or armchair) configuration [39,40]. In this regard, the absorption spectra in Figure 10 were simulated with respect to an impulse excitation polarized in the $x$-axis and $y$-axis directions, where the x-axis corresponded to the armchair edge direction and the y-axis was referred to a zigzag edge. Obviously that changing the edge direction leads to the pronounced alterations in the optical absorption spectra of all simulated systems, which are especially manifested below 4 eV. When the impulse excitation is polarized in the armchair edge direction ($x$ axis), the absorption peaks of HPSi are mainly located in the vicinity of the energy resonance points at 0.86, 1.51, and 2.53 eV; the absorption peaks of HSiP arise at 1.02, 1.53, 2.23, 2.88, and 3.89 eV, and those of HSiPbp are observed at 0.41, 1.37 and 2.62 eV. In the case of the impulse excitation polarized in the zigzag edge direction ($y$ axis), the absorption peaks of HPSi emerge at 0.58, 1.16 and 1.98 eV; those of HSiP appear at 0.94, 1.48, 2.59, and 3.24 eV, and the absorption signatures of HSiPbp are found at 0.96, 1.64, 2.46 and 3.1 eV. The energy resonance points which are below 1.6 eV refer to IR radiation. Those that are between 1.6 eV and 3.2 eV refer to visible light. Those that are higher than 3.2 eV refer to ultraviolet light

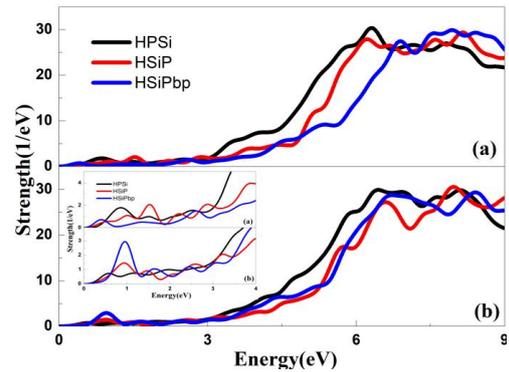

FIG. 10. The absorption spectra of HPSi, HSiP, and HSiPbp monolayer nanostructures to an impulse excitation polarized in the (a) $x$ axis and (b) $y$ axis directions below 6 eV. The inset in FIG. 10(b) shows a close-up view of the absorption spectra below 4 eV.

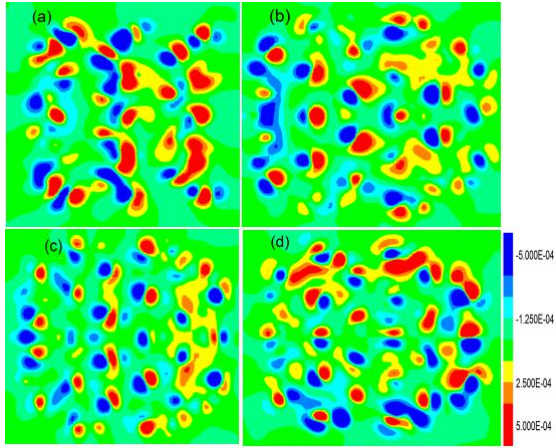

FIG. 11. Fourier transforms of the induced charge density of (a) HSiPbp and (b-d) HSiP nanostructures. The polarization direction is set along the armchair edge at energy resonance points of (a) 1.37 eV, (b) 2.23 eV, and (c) 2.88 eV and along the zigzag edge direction at energy resonance point of (d) 1.48 eV.

To elucidate the mechanism governing the behavior observed in the optical absorption spectra within the low-energy range for all the simulated structures, the spatial dependence of the induced charge response was analyzed at the resonance frequencies obtained from the time evolution in a plane. For this, the induced charge plane was assumed to be parallel to the graphene plane and the vertical distance between the top atomic layer of the nanostructure and the induced charge density plane was estimated to be 0.9 Å. The low-energy resonances were suggested to be localized at the boundary region. The induced charge density profiles for these plasmonic resonance points exhibited a dipole-like character. Moreover, the lower-energy plasmons follow the long-range charge transfer plasmon (CTP) mode attributed to the electronic motion along the direction in which electrons can propagate through longer distances. Figure 11 shows the Fourier transforms of the induced charge densities of HSiPbp and HSiP nanostructures. The polarization direction was set along the armchair edge at energy resonance points of 1.37 eV for HSiPbp nanostructure and at 2.23 and 2.88 eV for HSiP monolayer, as well along the zigzag edge direction at energy resonance points of 1.48 eV for HSiP monolayer. Compared with periodic systems, the selected nanostructures display the highly tunable polarization-dependent enhanced plasmon resonance in a wide frequency region, which would be useful for water splitting in infrared and visible light regions.

4. Conclusions

In summary, monolayer HPSi, HSiP, and HSiPbp were theoretically studied based on first-principles calculations. The calculations of the phonon spectra and cohesive energies revealed the high stability of the monolayer H-Si-P structures. According to the bandgaps and band edge simulated using the accurate hybrid functional, all the three monolayer structures belong to indirect bandgap semiconductors that were shown to be suitable for photocalitic water splitting under visible (HPSi) and ultraviolet light (HSiP and and HSiPbp). Furthermore, their bandgaps and band edge positions were found to be effectively adjusted by applying a biaxial tensile or compressive strain. Finally, the ability to use the simulated H-Si-P monolayers for light-induced photocatalytic water splitting was studied via the modeling of their optical absorption spectra without strain and under the ultimate biaxial tensile strain. The absence of absorption peaks within a visible-light range led to a need to calculate the absorption spectra of these nanostructures using a TDDFT method, which exhibited a series of absorption splitting within a visible range. In this respect, the simulated 2D nanostructures were shown to have excellent potential for solar energy conversion and visible-light-driven photocatalytic water splitting. Thus, this work provides valuable guidance for discovering the new 2D semiconductors and nanostructures for nanoelectronic and optoelectronic devices, as well as for potential photocatalytic water splitting applications.

Acknowledgments

Xiaoqin Shu and Jiahe Lin contribute equally to the article, Xiaoqin Shu and Jiahe Lin are co-first authors of the article. Hong Zhang acknowledges financial support from the National Natural Science Foundation of China (Grants No. 11474207) and National Key R&D Program of China (2017YFA0303600); Xiaoqin Shu acknowledges scientific research project of Leshan Normal University (No. XJR17007, and LZDP012 and DGZZ202009 ) , Key Research Project of Leshan Science Technology Bureau (No.20GZD036). Jiahe Lin acknowledges the Foundation from Department of Science and Technology of Fujian Province (China, Grant No. 2019L3008) , the Foundation from Department of Science and Technology of Fujian Province (China, Grant No. 2020J05147).